\documentstyle [12pt,psfig] {article}

 1
\def\Onot{\Omega_0}
\def\rm{\mathrm}
\def\erfc{\rm {erfc}}
\def\bbf{\mathbf}
\def\mpc{\,{\rm {Mpc}}}
\def\mpch{\,h^{-1}{\rm {Mpc}}}
\def\kms{\,{\rm {km\, s^{-1}}}}
\def\xicc{ { \xi}_{\rm {cc}}}
\def\xiN{ { \xi}_{\rm {N}}}
\def\msun{{M_\odot}}
\def\sigmaba{\sigma_8}
\def\xim{\xi_{\rm {m}}}

\def\deltac{\delta_c}
\def\ref{\parskip=0pt\par\noindent\hangindent\parindent
    \parskip =2ex plus .5ex minus .1ex}
\def\gs{\mathrel{\raise1.16pt\hbox{$>$}\kern-7.0pt
\lower3.06pt\hbox{{$\scriptstyle \sim$}}}}
\def\ls{\mathrel{\raise1.16pt\hbox{$<$}\kern-7.0pt
\lower3.06pt\hbox{{$\scriptstyle \sim$}}}}
\def\gtsima{$\; \buildrel > \over \sim \;$}
\def\ltsima{$\; \buildrel < \over \sim \;$}
\def\prosima{$\; \buildrel \propto \over \sim \;$}
\def\gsim{\lower.5ex\hbox{\gtsima}}
\def\lsim{\lower.5ex\hbox{\ltsima}}
\def\simgt{\lower.5ex\hbox{\gtsima}}
\def\simlt{\lower.5ex\hbox{\ltsima}}
\def\simpr{\lower.5ex\hbox{\prosima}}
\def\la{\lsim}
\def\ga{\gsim}
\begin{document}
\topskip=5.0truecm
\baselineskip=20pt
\centerline {\bf {THE CORRELATION FUNCTION OF CLUSTERS OF GALAXIES} }
\smallskip
\centerline {\bf {AND THE AMPLITUDE OF MASS FLUCTUATIONS IN THE UNIVERSE}}
\smallskip 
\vskip20pt
\centerline {H.J. Mo, Y.P. Jing, S.D.M. White}
\vskip10pt
\centerline {Max-Planck-Institut f\"ur Astrophysik}
\centerline {Karl-Schwarzschild-Strasse 1}
\centerline {85748 Garching, Germany}

\vskip130pt
\centerline {MNRAS: submitted}
\bigskip
\bigskip
\vfill
\eject

\topskip=0.0truecm
\centerline{\bf {ABSTRACT}}
\smallskip

We show that if a sample of galaxy clusters is complete above some
mass threshold, then hierarchical clustering theories for structure
formation predict its autocorrelation function to be determined purely
by the cluster abundance and by the spectrum of linear density
fluctuations. Thus if the shape of the initial fluctuation spectrum is
known, its amplitude $\sigma_8$ can be estimated directly from the
correlation length of a cluster sample in a way which is independent
of the value of $\Omega_0$. If the cluster mass corresponding to the
sample threshold is also known, it provides an independent estimate of
the quantity $\sigmaba\Onot^{0.6}$. Thus cluster data should allow
both $\sigmaba$ and $\Onot$ to be determined observationally. We
explore these questions using N-body simulations together with a
simple but accurate analytical model based on extensions of
Press-Schechter theory. Applying our results to currently available
data we find that if the linear fluctuation spectrum has a shape
similar to that suggested by the APM galaxy survey, then a correlation
length $r_0$ in excess of $20\mpch$ for Abell clusters would require
$\sigmaba>1$, while $r_0<15\mpch$ would require $\sigmaba<0.5$. With
conventional estimates of the relevant mass threshold these imply
$\Onot\la 0.3$ and $\Onot\ga 1$ respectively.


\bigskip
\centerline {\bf {1. INTRODUCTION}}
\bigskip

 Observations have shown that clusters of galaxies are strongly
clustered in space. The cluster-cluster two-point correlation
function, $\xicc (r)$, is roughly a power law,
$\xicc (r) =(r_0/r)^{\alpha}$, with $\alpha \sim 1.8$ and with
a correlation length $r_0$ much larger than that of galaxies
(see Bahcall 1988 for a review). The exact value of $r_0$ is,
however, still controversial. Redshift surveys of clusters
from the Abell catalogue (Abell 1958) and its southern
extension (Abell, Corwin \& Olowin 1989, ACO) give $r_0\sim
20$-$25\mpch$ \footnote {Throughout this paper, we write
the Hubble constant as $H_0=100h\kms \mpc ^{-1}$}
(Bahcall \& Soneira 1983; Klypin \& Kopylov 1983; Postman, Huchra \&
Geller 1992; Peacock \& West 1992). These data also show that the
correlation length $r_0$ increases with cluster richness ${\cal{R}}$
according to 
a scaling relation $r_0\approx 0.4 d$, where $d$ is the mean
intercluster separation and is related to the mean space density
$n$ by $d=n^{-1/3}$ (Bahcall \& Burgett 1986; Bahcall \& West
1992). The reliability of these results has, however,
been questioned by a number of authors (e.g. Sutherland 1988;
Dekel et al. 1989; Efstathiou et al. 1992b) who argued that they may
be affected by projection effects intrinsic to the construction of the
Abell/ACO catalogues. These authors claim that the
correlation function of ${\cal {R}}\ge 1$ Abell clusters may have been
significantly overestimated. Other authors (e.g.
Bahcall \& West 1992; Jing, Plionis \& Valdarnini 1992;
Peacock \& West 1992) claim that 
any bias in $\xicc$ due to projection effects 
is small. Recent estimates of $r_0$ based on the APM
cluster catalogue (Dalton et al. 1992, 1994) and the Edinburgh-Durham
cluster catalogue (Nichol et al. 1992) give $r_0=13-16\mpch$
for clusters with $d\approx 35-45\mpch$ (compared to 
$d\approx 55\mpch$ for ${\cal {R}}\ge 1$ Abell clusters). These results
are still consistent with the $r_0$-$d$ relation 
described above. Since these catalogues are too small
to give an accurate estimate of $r_0$ for richer clusters, 
the controversy over the correlation length of rich Abell
clusters is still unresolved.

  Despite the uncertainties in current observational results,
the correlation function of clusters of galaxies remains 
one of the most important diagnostics for models of structure
formation. Indeed, the strong observed correlation of clusters on large 
scales is very difficult to 
reconcile with the standard cold dark matter (CDM) model
(White et al. 1987; Dalton et al. 1992; Jing et al. 1993; 
Mo, Peacock \& Xia 1993; Dalton et al. 1994). In the near future,
as new data from x-ray observations and large digital sky
surveys become available, cluster correlation functions 
will provide much more stringent constraints on models.
In addition, future observations of cluster x-ray properties and
of gravitational lensing by  clusters will provide 
improved estimates for the masses of clusters. We can then 
examine $\xicc$ as an explicit function of 
cluster mass.

It is clearly important to understand how cosmological models are
constrained by such observations. In this context, a simple analytic
model for $\xicc$ and for cluster abundances is particularly desirable.

  The observed high amplitude of $\xicc$ is usually considered 
to be a result of clusters being high peaks in the initial
density field (Kaiser 1984). Although qualitatively correct, this
peak theory does not provide an adequate estimate for the two-point
correlation functions of clusters in N-body simulations 
(Croft \& Efstathiou 1994; Mann, Heavens \& Peacock 1993).
Furthermore, it is not clear how to derive a mass function
for clusters from the peak theory, because peaks of small
masses may be contained by those of larger masses.

An alternative scheme, the Press-Schechter formalism (Press \&
Schechter 1974, PS) defines objects (halos) to be virialized
structures, identified at the present time,  which have grown from
Gaussian initial density fluctuations.  Although the original
derivation of PS is flawed and recent rederivations are still far from
rigorous (Bower 1991; Bond et al.  1991), the PS formalism gives
surprisingly good fits to the mass functions of dark matter halos
identified in N-body simulations of hierarchical clustering
(Efstathiou et al. 1988; White, Efstathiou \& Frenk 1993; Lacey \&
Cole 1994).  Recently, Mo \& White (1995) show that the PS formalism
and its extensions can also be used to derive a model for the spatial
correlation of dark matter halos in hierarchical models. They find
that the halo-halo two-point correlation functions predicted by this
model agree surprisingly well with those derived from N-body
simulations of evolution from scale-free initial power spectra. Such a
model can obviously help us to understand better the clustering
properties of clusters of galaxies. A related model has been used by
Kashlinsky (1987) to explain the richness dependence of cluster
correlations.

  In this paper, we present additional tests of the model
of Mo \& White for realistic initial power spectra
and in the regime relevant to clusters of galaxies. We also
discuss how these results can be used to relate observations
of the large-scale clustering of clusters to the fundamental
parameters of cosmology.
We describe our model in Section 2. The N-body simulations
used to test our model are described in Section 3. 
Comparisons of model predictions with simulation results
are presented in Section 4. In Section 5 we apply our 
model to real clusters. A brief discussion of our results
is given in Section 6.

\bigskip
\centerline {\bf {2. THE MODEL}}
\bigskip

 In this paper, we consider models in which the universe is spatially
flat (so that $\Omega_0+\Omega_\Lambda=1$, where $\Omega_0$ and
$\Omega_\Lambda$ are cosmic density parameters referring to matter and
to the cosmological constant respectively) and dominated by cold dark
matter. We also assume that the primordial density field is Gaussian,
with a scale-invariant power spectrum. Neglecting the contribution of
baryonic matter to $\Omega_0$, the initial power spectrum of density
fluctuations can be written as (see Bardeen et al. 1986, equation G3):
$$
P(k)\propto {k T^2(k)},
\eqno(1a)
$$
with
$$
T(k)={ \ln(1+2.34q)\over 2.34 q}
\left\lbrack 1+3.89 q+(16.1q)^2+(5.46q)^3+(6.71q)^4
\right\rbrack^{-1/4}
\eqno(1b)
$$
and
$$
q\equiv {k\over \Gamma h {\mpc}^{-1}} ,
\eqno(1c)
$$
where, following Efstathiou et al. (1992a), we have
introduced a shape parameter, $\Gamma\equiv \Omega_0 h$,
for the power spectrum.

 A useful function equivalent to the power spectrum is 
the rms mass fluctuation in a spherical top-hat window of radius 
$R$:
$$
\sigma ^2(R)=
\sum_{{\bbf k}} P(k) {\hat W}^2(R; k),
\eqno(2)
$$
where ${\hat W}(R, k)$ is the Fourier transform
of the top-hat window. We normalize the power spectrum
by specifying $\sigmaba\equiv \sigma(8\mpch)$. We define
a mass parameter $M_8$ as the mean mass 
contained in a sphere with radius $8\mpch$ in a universe
with the critical density, thus
$M_8=5.8\times 10^{14} h^{-1} \msun$.
When lengths are written in units of $\mpch$ and masses 
in units of $M_8$, all quantities we are interested in are
independent of the Hubble constant. A model will then be
specified by giving ($\Omega_0,\Gamma,\sigmaba$).

In the PS formalism, dark matter halos are defined as 
spherically symmetric, virialized clumps of dark matter
particles. The mass $M$ of a halo is related to 
its initial comoving  radius $R$ (measured in current units)
by
$$
M={4\pi \over 3} {\bar \rho}R^3,
\eqno(3)
$$
where ${\bar \rho}$ is the mean mass density of the universe
at present time. Thus the rms mass fluctuation on the
scale of the halo is $\sigma (R)$.
The fraction of matter in halos with mass exceeding $M$ is
$$
{\cal F} (\nu )={\erfc} \left\lbrack {\nu \over \sqrt {2}}
\right\rbrack ,
\eqno (4)
$$
where $\nu \equiv \deltac/ \sigma(R)$ and ${\erfc} (x)$ is 
the complementary error function. The critical linear
overdensity for virialization, $\deltac$, is chosen to be
$\deltac=1.69$, irrespective of $\Omega_0$, as discussed in
White et al. (1993, hereafter WEF). The PS formula for the comoving 
number of halos with mass $M$ is
$$
n(M){\rm d}M =
-\left({2\over \pi}\right)^{1/2} {{\bar \rho}\over M}
{\deltac \over \sigma}
{{{\rm d\,}\ln}\sigma \over {\rm d}\,\ln M}
{\exp} \left\lbrack -{\deltac^2 
\over 2 \sigma ^2 }\right\rbrack 
{{\rm d} M \over M}.
\eqno(5)
$$
It follows that the total number density of halos with mass 
exceeding $M$ is
$$
n(>M)=-{3\over (2\pi)^{3/2}}\int_R^\infty
{1\over x^3}{\deltac\over \sigma (x)}
{{\rm d\,}\ln\sigma \over {\rm d\,}\ln x}
{\exp} \left\lbrack -{\deltac^2 
\over 2 \sigma ^2 }\right\rbrack 
{{\rm d} x \over x}.
\eqno(6)
$$
Thus for given $\sigma(R)$ the abundance of halos depends only on
$R$. An $\Omega_0$-dependence enters only through the relation between
$R$ and $M$. 

 According to Mo \& White (1995), the two-point correlation function
of dark matter halos with mass $M$ is related to that
of the mass, $\xi_{\rm m}$, by
$$
\xi (r)=
\left\lbrack b(M)\right\rbrack^2\xi _{\rm m}(r),
\eqno(7a)
$$
where
$$b(M)=1+{1\over \deltac}(\nu^2-1).
\eqno(7b)
$$ 
This is similar to what one gets from the peak-background split argument
(Efstathiou et al. 1988; Cole \& Kaiser 1989).
When halos with a range of masses are considered,
$b(M)$ in equation (7b) should be replaced by its 
average [weighted by $n(M)$] over $M$.
Using N-body simulations of scale-free spectra, Mo \& White
show that equation (7) holds even for $\xi_{\rm m}(r)\ga 1$, as long as
$r$ is not smaller than the Lagrangian radius $R$ of halos.
For rich clusters of galaxies, this Lagrangian radius is
$R\sim 10\mpch$ (WEF).
Thus we expect equation (7) to be valid for 
$\xicc (r)$ on $r\gsim 10\mpch$. On these scales, $\xi_{\rm m}(r)$ can
be represented sufficiently accurately for our purposes
by the Fourier transform of the
linear power spectrum. For a given shape
of power spectrum, the abundance of clusters gives a relation between
the threshold $R$ and the amplitude $\sigma_8$ (through equation
6). The autocorrelation function gives another such relation through
the definition $\nu\equiv \delta_c/\sigma(R)$ and equation (7). 
These two relations determine both $R$ and
$\sigmaba$. An estimate of the cluster mass at threshold then
determines $\Onot$
  
\bigskip
\centerline {\bf {3. N-BODY SIMULATIONS}}
\bigskip

We use four sets of ${\rm P}^3{\rm M}$ N-body simulations to test the
theories presented in the previous section. Each simulation can be
characterized by three model parameters ($\Omega_0, \Gamma, \sigmaba$)
(as discussed in Section 2), and three simulation parameters:
the box size $L$ (in $\mpch$), the number of simulation particles $N$ and the
effective force resolution $\eta$ (in $\mpch$) . 
For the first set of simulations, we
use $(\Omega_0, \Gamma, \sigmaba)=(0.3,0.225,1)$ and 
$(L,N,\eta)=(400,100^3,0.2)$. Six realizations were run
for this set. 
The second set consists of five realizations of the SCDM universe with
$(\Omega_0, \Gamma, \sigma_8)=(1.0, 0.5, 1.24)$ and
$(L,N,\eta)=(300, 128^3, 0.23)$.
The third is a single realization with ($\Omega_0, \Gamma)
=(1,0.2)$ and $(L,N,\eta)=(256,128^3,0.2)$. 
Analysis is made for two output times with
$\sigmaba=1$ and $\sigmaba= 0.5$.
The final set is another single simulation with the same 
initial density spectrum 
(i.e. with $\Gamma=0.2$) and the same
simulation parameters as the third set, 
but for a low density flat universe with 
$\Omega_0=0.2$. The two output times for this case
correspond to $\sigma_8=1.07$ and 0.5, respectively.
The mass of each individual particle
is $5.3\times 10^{12}\msun$, $3.6\times 10^{12}\msun$, 
$2.2\times 10^{12}\msun$
and $4.5\times 10^{11}\msun$ 
in these four sets of simulations.

\bigskip
\centerline {\bf {4. COMPARISON WITH N-BODY SIMULATIONS}}
\bigskip
  For most of our discussion, we will use clusters identified
by the standard friends-of-friends (FOF) group finder. 
A cluster in a simulation
is then defined as all mass particles that are connected by joining
particle pairs with separations smaller than a given linkage
length, $l$, in units of the mean interparticle separation. 
There is no {\it a priori} reason for a particular
choice of $l$. Following common practice, we choose $l=0.2$
(see e.g. Davis et al. 1985; Efstathiou et al. 1988; 
Lacey \& Cole 1994). The mean
density within a cluster (${\bar \delta}$) is then of the order
200, in rough agreement with what one obtains for a spherically
symmetric, virialized object in an Einstein-de Sitter universe
(Gunn \& Gott 1972). For a spatially flat universe with $\Omega_0=0.2$,
${\bar \delta}$ predicted by the virialization model is about
$2.5$ times higher (WEF), and a smaller linkage length
might seem more reasonable. However, the real virialization process
must be much more complicated than a spherical accretion model, and
the appropriate value of $l$ may depend on cluster density profiles, which
are different in different models (e.g. Jing et al. 1995). 
In the following we will see
that our choice of $l$ gives a mass function that is 
in good agreement
with that predicted by the PS formalism, even for a low-density
flat universe. For comparison, we will also present 
some results obtained using the spherical-overdensity (SO) grouping
algorithm invented by Lacey \& Cole. 
This algorithm is based on finding spherical regions with 
a certain predefined mean overdensity $\kappa$. A local density near 
each particle is needed to provide an initial list of possible halo
centres, and is defined as $3(N+1)/(4\pi r_N^3)$, where $r_N$ is the 
distance to the N'th nearest neighbour. Further details may be found
in Lacey \& Cole (1994). We follow them in choosing $\kappa=180$ and 
$N=10$. In this algorithm the mass of a cluster is simply the number of 
particles within the bounding sphere.

\vfill
\eject
\bigskip
\centerline {\bf {4.1. Cluster number densities}}
\smallskip

In Figure 1 the cumulative number densities of clusters 
in the 
simulations (circles) are compared directly with the predictions
of the PS formalism (dashed curves). The error bars in Fig.1a and 1b
represent the $1\sigma$ standard deviations between different 
realizations,
whereas those in Fig.1c and 1d represent $1\sigma$ Poisson 
fluctuations of the cumulative number density. This figure shows
that the model prediction agrees with the simulation results reasonably
well over a wide range of masses. The abundance range relevant for
clusters of galaxies is $n(>M)=10^{-4.5}$-$10^{-6} (\mpch)^{-3}$.
The model works extremely well for these abundances. This is
a nontrivial result,
because the same $\deltac$ and $l$ have been used for all cases. We have
tried using $l=0.15$ for models with $\Omega_0=0.2$, and found that
the agreement between the resulting mass function and model prediction
is considerably worse. 
For comparison, we also show in Fig.1c and 1d the results for clusters
identified by the SO group finder (crosses). It is clear that the 
mass function given by this group finder is similar to that given by
the FOF group finder with linkage radius $l=0.2$. This result
is in agreement with that obtained by Lacey \& Cole (1994) for
scale-free initial density spectra evolved in an Einstein-de Sitter
universe.
 
\bigskip
\centerline {\bf {4.2. The two-point correlation function}}
\smallskip
  The two-point correlation function of clusters in the 
  simulations is estimated directly from pair counts:
  $$
  \xicc(r)={\Delta P (r)\over 4\pi nr^2\Delta r} -1,
  \eqno(8)
  $$
  where $\Delta P(r)$ is the average number of neighbors,
  per cluster, with separation in the range $r\pm \Delta r/2$ and
  $n$ is the mean number density of clusters in the sample.
  The circles in Figures 2-7 show $\xicc$ for the FOF clusters
  identified in the N-body
  simulations. Results are shown for clusters in different mass bins,
  as indicated in the panels by the number of mass particles, $N_p$.
  The corresponding masses of clusters can be obtained by 
  multiplying by the mass of each individual particle.
  The error bars in Figs.2 and 3 
  represent the $1\sigma$ scatter between realizations,
  while those in Figs.4-7 represents a $1\sigma$ error based on 
  Poisson statistics: $\Delta \xicc (r)=[\Delta P(r)]^{1/2}/
  [4\pi nr^2\Delta r]$. For comparison, we show as solid curves
  the fits of the data points at $r>10\mpch$ to a model,
  $\xicc(r)=A\xim(r)$, where $A$ is constant. 
In this fitting, each data point is weighted by the 
  inverse
  of its bootstrap error, as is discussed in Mo, Jing \& B\"orner 
  (1992). The crosses in Figs.4-7 show $\xicc$ for SO clusters
  with the same mass as the corresponding FOF clusters shown in the
  same panel. In agreement with the results obtained by 
  Mo \& White (1995) for scale-free initial density spectra, the
  correlation functions for FOF and SO clusters of the same 
  mass are quite similar. 
  The dashed curves in Figs.2-7 show the predictions of equation (7).
 The figures show that theory 
  and simulations agree remarkably well for all cases.

  \bigskip
  \centerline {\bf {5. APPLICATION TO REAL CLUSTERS}}
  \bigskip
  As shown in Section 2, for a given shape of power spectrum
  the autocorrelation function and the number density
  of clusters with Lagrangian scale exceeding $R$ are both determined 
  purely by $\sigmaba$. 
  A comparison of the two allows $R$ to be eliminated, determining
  $\sigmaba$ in a way which is independent of $\Onot$.

  Figure 8 shows the relation between the correlation length $r_0$ and
the intercluster separation $d$ for two sets of linear power
  spectra. One has $\Gamma=0.5$ (Fig.8a), as in the standard 
CDM model, the other has $\Gamma=0.2$ (Fig.8b),
  which is similar to the shape obtained from the angular two-point
  correlation function of galaxies in the APM survey
  (see Efstathiou, Sutherland \& Maddox, 1990). 
  It is clear that a stringent 
  constraint on $\sigmaba$ can be obtained from an accurate
  measurement of $r_0$.
  The squares are observational results 
  for different cluster samples as compiled by Bahcall \& Cen (1992),
while the cross shows the result of a recent analysis 
 by Dalton et al. (1994). 
  The data point at $d=94\mpch$ is based on Abell ${\cal {R}}\ge 2$ 
  clusters, and is very uncertain due to the small size of the
  sample (Peacock \& West 1992).
  As one can see from Fig.8a, the SCDM model with $\sigmaba=0.5$
  does not have enough large-scale power to match the observational
  data. Indeed, if $r_0\ge 20\mpch$ for clusters with
  $d\sim 55\mpch$, then SCDM models with any reasonable $\sigmaba$ can
  be ruled out. In contrast, models with $\Gamma =0.2$ 
  are consistent with such correlation lengths provided 
  $\sigmaba\gsim 1$ (Fig.8b). As one can see from Fig.8b, in order
  to obtain an accurate value of $\sigmaba$, it is crucial to
  measure $r_0$ accurately for such rich clusters. If $r_0$ for 
  clusters with $d\approx 55\mpch$ 
  is actually significantly overestimated and its true value is
  smaller than $15\mpch$ (Sutherland 1988; Efstathiou et al. 1992b), 
  then $\sigmaba<0.5$ is required  
  for the power spectrum shape inferred from the APM galaxy sample. 
  At present, the observational data are too uncertain to  
  provide reliable constraints. In the future, as new data
  from digital sky surveys and x-ray observations become available, 
  these arguments should provide a determination of $\sigmaba$ which
  can be compared directly with the
  amplitude on larger scales inferred from
  the COBE measurements of cosmic microwave background anisotropy
  (Smoot et al. 1992; Wright et al. 1994).

  Based on the observed masses and abundances of rich clusters of 
  galaxies, WEF obtained the constraint $\sigmaba\Omega_{0}^{0.56}
  \approx 0.57$,
  for a spatially flat universe. A similar constraint can be 
  obtained from the observed masses and correlation functions.
  As discussed in WEF, the Lagrangian radius of
  rich clusters, $R$, is about $r_8\equiv 8\mpch$. For $R$ near 
  $r_8$ the rms mass fluctuation, $\sigma (R)$, can be approximated
  by 
  $$\sigma (R) \approx \sigmaba(r_8/R)^{\gamma},
  \eqno(9)
  $$
  where the index $\gamma$ measures the local slope of the 
  fluctuation spectrum, and is given by $\gamma=0.68+0.4\Gamma$
  for the CDM-like spectra discussed here (Efstathiou, Bond \&
  White, 1992a). Using equation (9) and 
  equation (7) we obtain
  $$
  \sqrt{{\xicc (r)\over \xiN (r)}}=\sigmaba
  \left(1-{1\over \deltac}\right)+\deltac
  \left\lbrack {M\over M_8}\right\rbrack ^{2\gamma/3}
  {1\over \sigmaba\Omega_0^{2\gamma/3}} ,
  \eqno(10)
  $$
  where $\xiN$ is the mass two-point correlation function
  for a linear power spectrum with $\sigmaba=1$;
  $M_8=5.8\times10^{14}h^{-1}\msun$, as introduced in Section 2. 
  For rich clusters, the first term in the r.h.s. of equation (10)
  can be neglected, and it follows that
  $$
  \sigmaba\Omega_0^{2\gamma/3}
  =\deltac\left\lbrack {M\over M_8}\right\rbrack^{2\gamma/3}
  \left\lbrack {\xicc \over \xiN}\right\rbrack ^{-1/2} . 
  \eqno(11)
  $$
  For rich clusters, $M\approx M_8$ (WEF)
  and $(\xicc/\xiN)\approx 10$. We get $\sigmaba\Omega_0^{0.5} 
  \approx 0.53$ for $\Gamma=0.2$. This result is very similar to  
  that obtained by WEF from cluster abundances. 
 
  Combining this result (or the result of WEF) with
  our result for $\sigmaba$, one can in principle get an estimate 
  for $\Omega_0$. Indeed, if the power spectrum of mass density 
  fluctuations has the shape suggested by the APM survey 
  (i.e. $\Gamma\approx 0.2$), 
  then $r_0\gsim 20\mpch$ for Abell clusters would require $\Omega_0 \lsim
0.3$, while $r_0 \lsim 15\mpch$ would require $\Omega_0 \gsim 1$. 
With better data on cluster masses and correlations, these arguments
should provide an estimate of $\Onot$.

\vfill
\eject
\bigskip
\centerline {\bf {6. DISCUSSION}}
\bigskip

Our simulations show that the mass function and the
autocorrelation function of clusters can be quite accurately 
predicted by our Press-Schechter model. As a result this model
provides a simple and intuitive way to understand how cluster
data constrain 
$\sigmaba$ and $\Omega_0$. A significant uncertainty remains, however,
in the operational definition of a galaxy cluster. 
In our model, as in most theoretical studies, clusters
are selected according to mass and overdensity thresholds.
The two group-finding algorithms we have tested give similar  
mass functions and two-point correlation functions.  
Unfortunately, the selection criteria in real 
observational samples can be far more complex
than assumed by these algorithms, and may differ from catalogue 
to catalogue. In a recent paper, Eke et al. (1995) used
N-body simulations of the SCDM cosmogony to examine how the two-point 
correlation function of a cluster sample depends on the way 
it is selected.
They found small but significant variations between cluster samples
defined in different ways.
Their results suggest that selection effects 
must be considered carefully, when a rigorous comparison 
between models and observations is made.
In the future, when cluster samples with well defined
selection criteria are available, one should use simulations
that take into account these selection criteria in as much detail 
as possible in order to calibrate the kind of measurements we are
suggesting. The observed masses and correlations of rich clusters
should then provide direct and independent measurements of the density
of the Universe and of the amplitude of mass fluctuations.
\vfill
\eject
\bigskip
\leftline {\bf {Acknowledgements:}} We thank S. Cole and
C. Lacey for providing us with their code for the SO group
finder. YPJ acknowledges the receipt of an 
Alexander-von-Humboldt research fellowship.
\vfill
\eject

\bigskip
\centerline {\bf REFERENCES}
\bigskip

\ref Abell G.O., 1958, {ApJS}, {3}, 211

\ref Abell G.O., Corwin H.G., Olowin R.P., 1989, ApJS,
{70}, 1

\ref Bahcall N.A., 1988, ARA\&A, {26}, 631

\ref Bahcall N. A., Burgett W., 1986, {ApJL},
{300}, L35

\ref Bahcall N. A., Cen R.Y., 1992, {ApJ},
{398}, L81

\ref Bahcall N. A., Soneira R. M., 1983, {ApJ},
{270}, 20

\ref Bahcall N. A., West M., 1992, {ApJ},
{392}, 419

\ref Bardeen J., Bond J.R., Kaiser N., Szalay A.S., 1986,
{ApJ}, {304}

\ref Bond J.R., Cole S., Efstathiou G., Kaiser N., 1991,
ApJ, 379, 440

\ref Cole S., Kaiser N., 1989, MNRAS, 237, 1127 

\ref Croft R.A.C., Efstathiou G., 1994, MNRAS, 267, 390

\ref Dalton G.B., Efstathiou, G., Maddox S.J., Sutherland W.J.,
1992, ApJ, 390, L1 

\ref Dalton G.B., Croft R.A.C., Efstathiou, G.,Sutherland W.J., 
Maddox S.J., Davis M.,
1994, MNRAS, 271, L47 

\ref Davis M., Efstathiou G., Frenk C.S., White S.D.M., 1985,
     ApJ, 292, 371

\ref Dekel A., Blumenthal G. R., Primack J. R., Olivier S.,
1989, ApJ, {338}, L5

\ref Efstathiou G., Bond J.R., White S.D.M., 1992a,
MNRAS, 258, 1P

\ref Efstathiou G., Dalton G.B., Sutherland W.J., Maddox S.J., 1992b,
MNRAS, 257, 125

\ref Efstathiou G., Frenk C.S., White S.D.M., Davis M.,
1988, MNRAS, 235, 715

\ref Efstathiou G., Sutherland W.J., Maddox S.J., 1990, Nat, 348, 705

\ref Eke V.R., Cole S., Frenk C.S., Navarro J.F., 1995, MNRAS, in press

\ref Gunn J. E., Gott J. R., 1972, {ApJ}, {176}, 1

\ref Jing Y.P., Mo H.J., B\"orner G., Fang L.Z., 1993, ApJ, 411, 450

\ref Jing Y.P., Mo H.J., B\"orner G., Fang L.Z., 1995, MNRAS, 276, 417

\ref Jing Y.P., Plionis M., Valdarnini R., 1992, ApJ, 389, 499

\ref Kaiser N., 1984, ApJL, 284, L9

\ref Kashlinsky M., 1987, {ApJ}, {317}, 19

\ref Klypin A., Kopylov A., 1983, {Sov. Astron. Lett.}, {9}, 41

\ref Lacey C., Cole S., 1994, MNRAS, 271, 676

\ref Mann R.G., Heavens A.F., Peacock J.A., 1993, MNRAS, 263, 798

\ref Mo H.J., Jing Y.P., B\"orner G., 1992, ApJ, 392, 452

\ref Mo H.J., Peacock J.A., Xia X.Y., 1993, MNRAS, 260, 121 

\ref Mo H.J., White S.D.M., 1995, preprint

\ref Nichol R.C., Collins C.A., Guzzo L., Lumsden S.L., 1992, MNRAS,
255, 21P

\ref Peacock J.A., West M.J., 1992, MNRAS, 259, 494

\ref Peebles P.J.E., 1980, The Large-Scale Structure of the Universe,
Princeton University Press, Princeton

\ref Postman M., Huchra J.P., Geller M.J., 1992, {ApJ}, {384},
404

\ref Press W.H., Schechter P., 1974, ApJ, 187, 425 (PS)

\ref Smoot G.F. et al. 1992, ApJ, 396, L1

\ref Sutherland W., 1988, MNRAS, {234}, 159

\ref White S.D.M., Frenk C.S., Efstathiou G., 1993, MNRAS, 262, 1023
(WEF)

\ref White S.D.M., Frenk C.S., Davis M., Efstathiou G., 1987, ApJ,
313, 505

\ref Wright E.L., 1994, ApJ, 420, 1
\vfill
\eject
\pagestyle{empty}

\begin{figure}
\centerline{
\psfig{figure=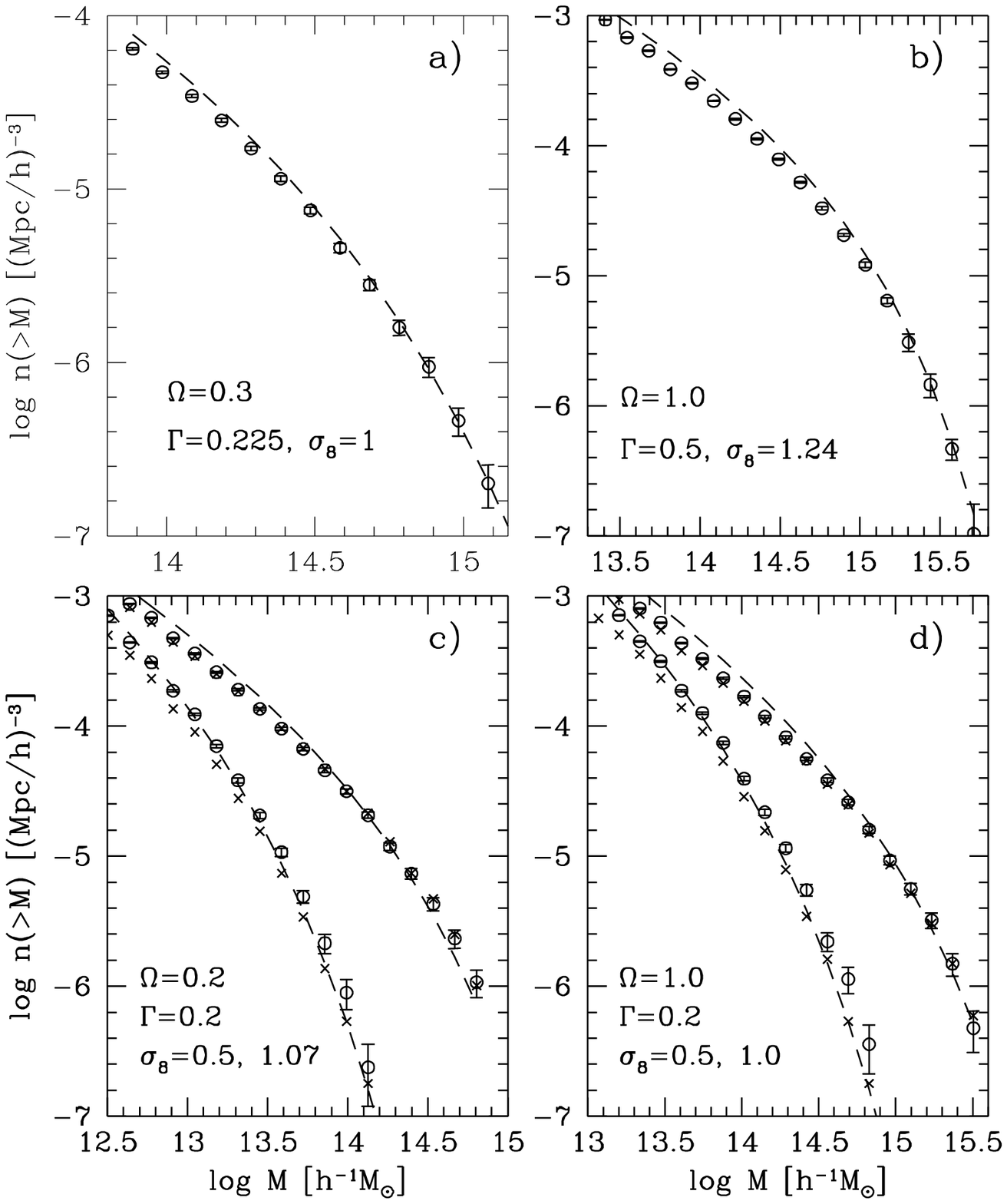,width=14.5cm,height=16.5cm}}
\noindent
{\bf Figure 1.} Cumulative mass function of FOF clusters
in N-body simulations (circles), compared to the predictions
of the PS formalism (dashed curves). Error bars in 
(a) and (b) denote the scatter between  different
realizations; those in (c) and (d) denote Poisson errors
in the number density. In (c) and (d), higher curves
correspond to higher values of $\sigmaba$. For comparison,
results for SO clusters are also ploted as crosses in (c) and (d).
\end{figure}

\begin{figure}
\centerline{
\psfig{figure=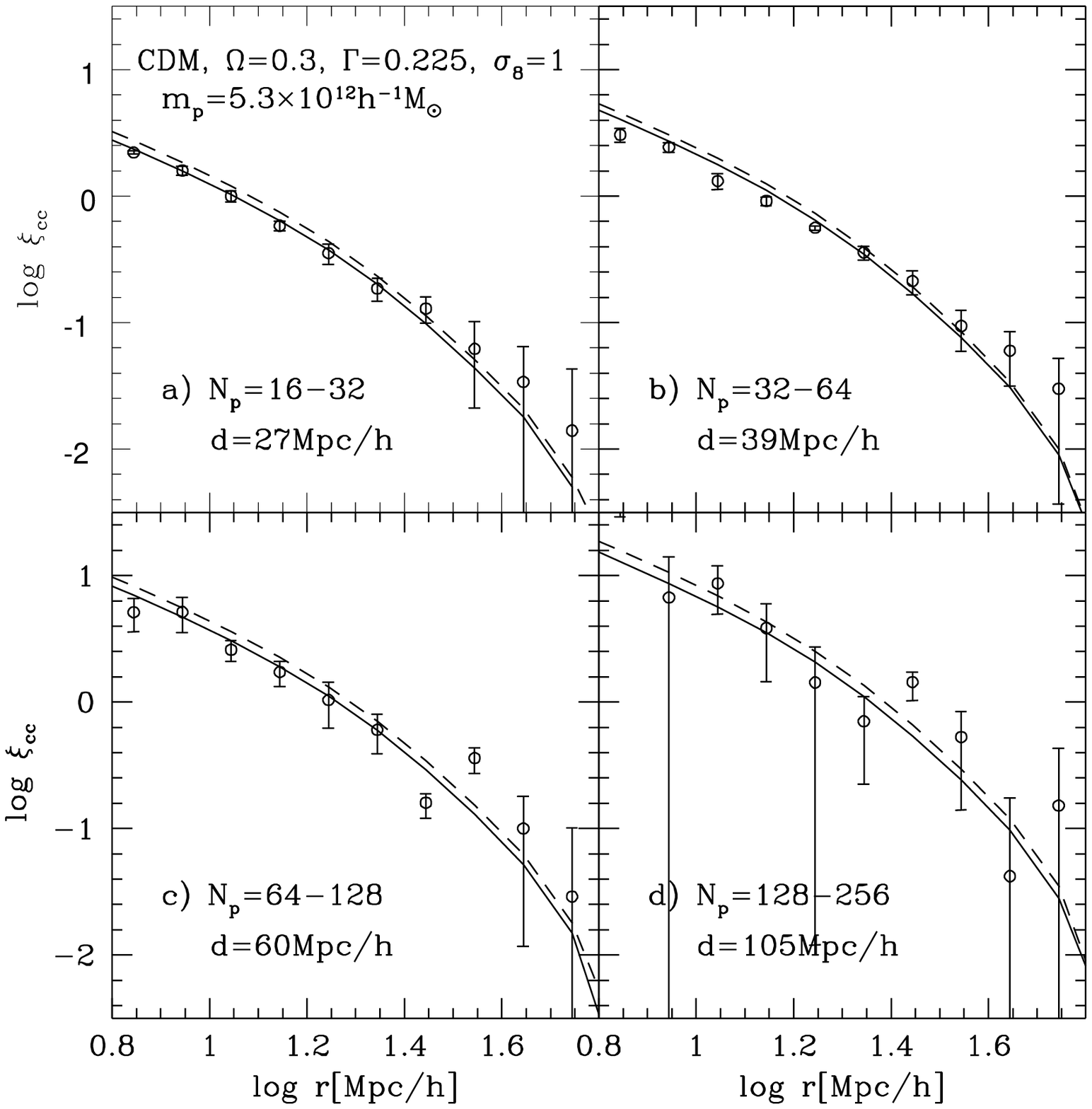,width=14.5cm,height=16.5cm}}
\noindent
{\bf Figure 2.} Two-point correlation functions of clusters
of galaxies identified by the FOF group finder in 
N-body simulations (circles), compared to 
model predictions (dashed curves). This is for a model
with $(\Omega_0, \Gamma, \sigmaba)=(0.3,0.225,1)$.
Solid curves show fits of the data points at $r>10\mpch$
to a model in which $\xicc(r)$ is proportional to the 
two-point correlation function corresponding to the linear power spectrum.
The error bars represent the $1\sigma$ scatter between 
different realizations. The value of $m_p$ denotes
the mass of a single mass particle in the simulation.
Results are shown for clusters in different mass ranges,
as indicated by the number of particles, $N_p$,
contained in these clusters. The value of $d$ denotes
the mean intercluster separation for clusters with masses
exceeding the lower limit of each mass bin.
\end{figure}

\begin{figure}
\centerline{
\psfig{figure=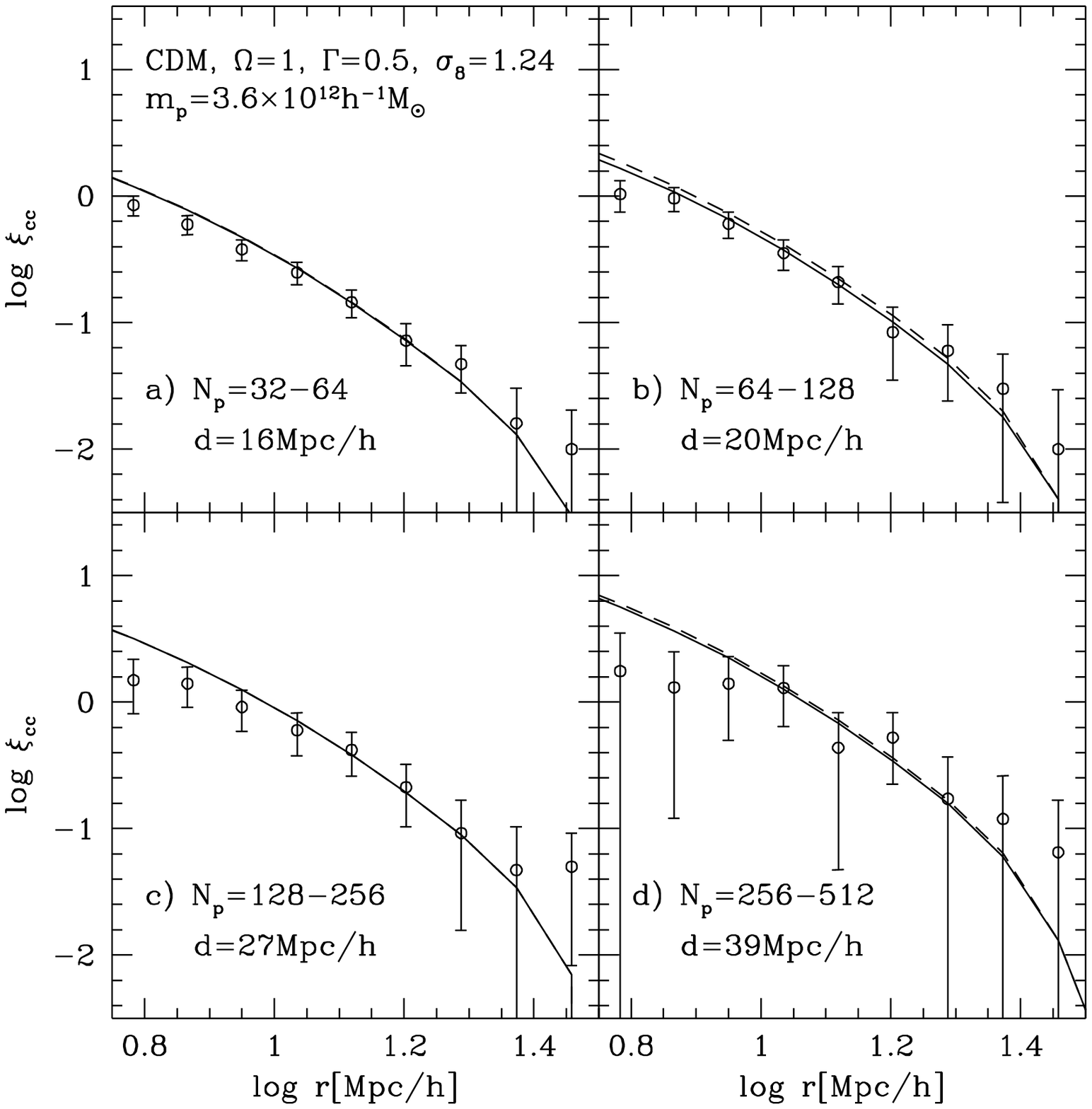,width=14.5cm,height=16.5cm}}
\noindent
{\bf Figure 3.} The same as Figure 2, for a model
with $(\Omega_0, \Gamma, \sigmaba)=(1,0.5,1.24)$.
\end{figure}

\begin{figure}
\centerline{
\psfig{figure=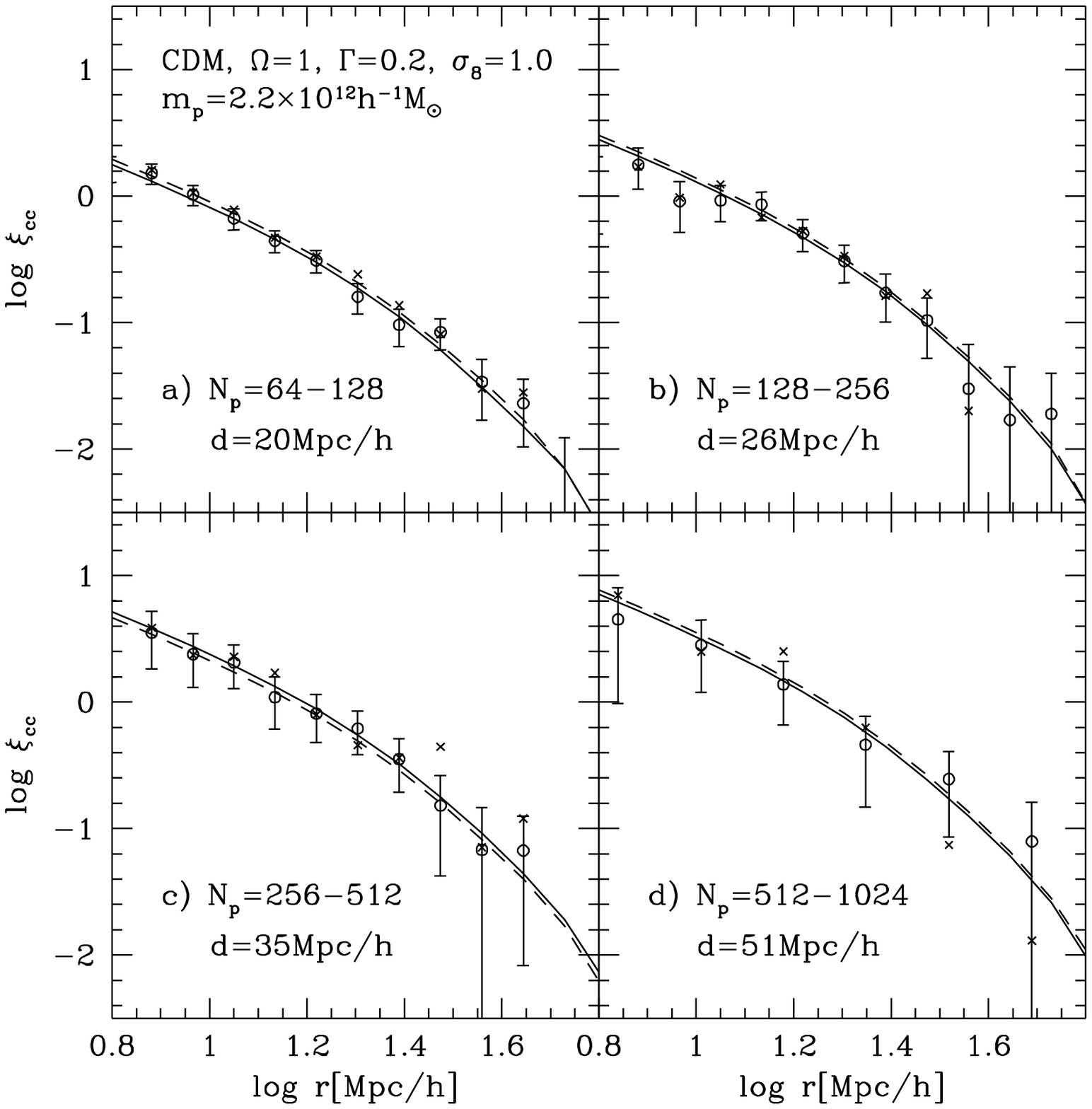,width=14.5cm,height=16.5cm}}
\noindent
{\bf Figure 4.} The same as Figure 2, for a model
with $(\Omega_0, \Gamma, \sigmaba)=(1, 0.2, 1)$.
Unlike in Fig.2, error bars here denote Poisson
fluctuations. 
Crosses show the same results for SO clusters.
\end{figure}

\begin{figure}
\centerline{
\psfig{figure=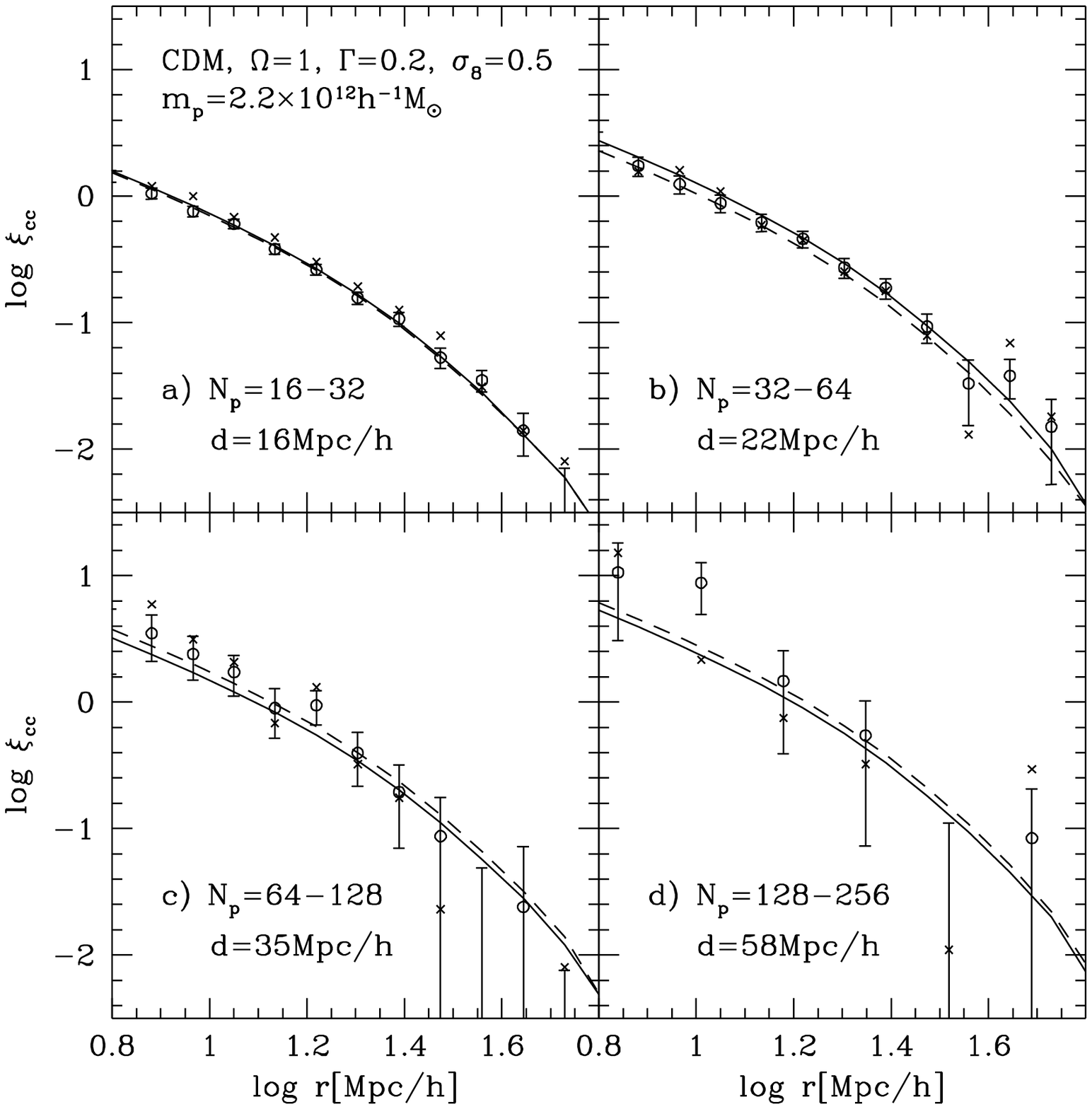,width=14.5cm,height=16.5cm}}
\noindent
{\bf Figure 5.} The same as Figure 4, for a model
with $(\Omega_0, \Gamma, \sigmaba)=(1,0.2,0.5)$.
Crosses show the same results for SO clusters.
\end{figure}

\begin{figure}
\centerline{
\psfig{figure=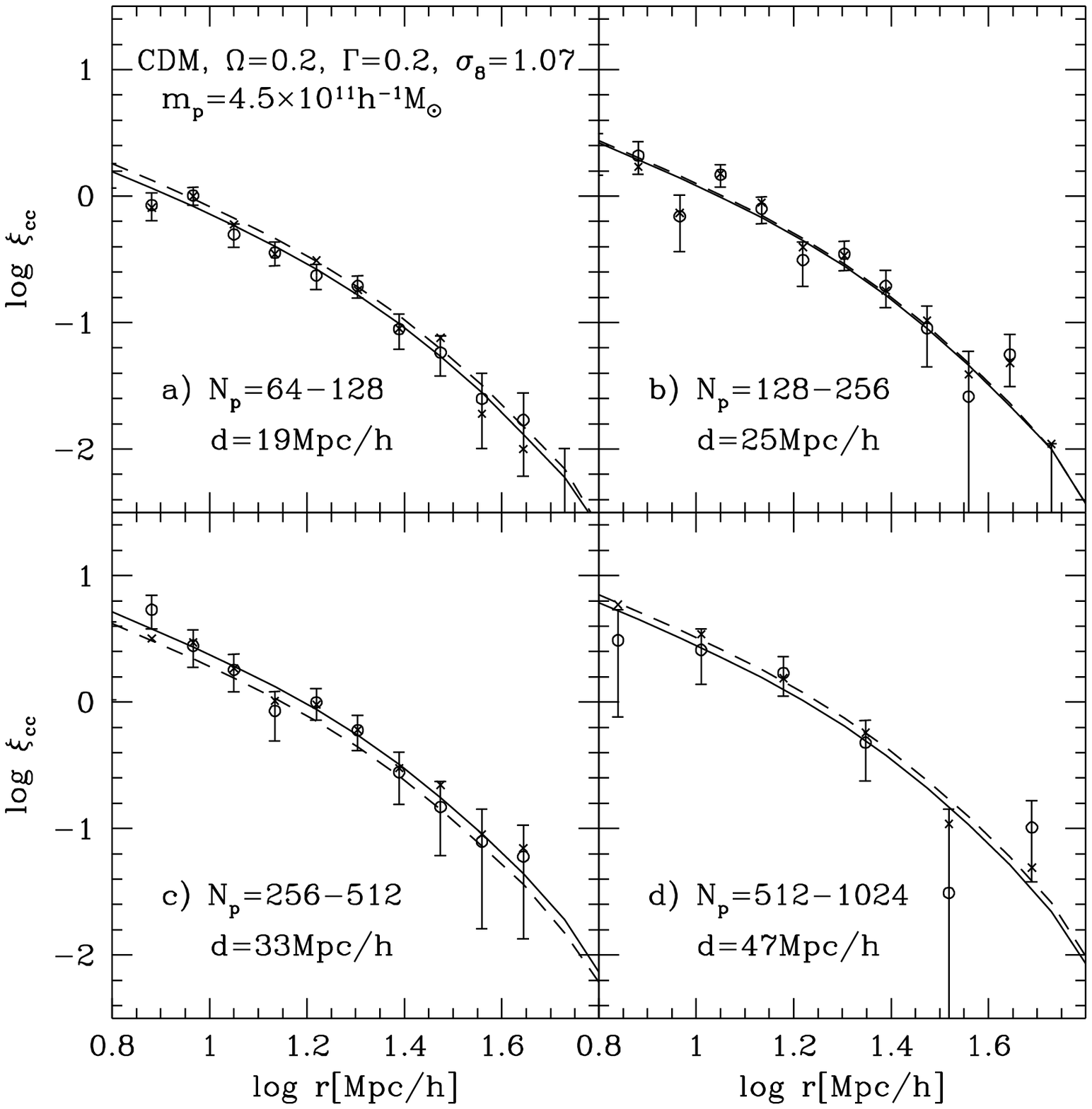,width=14.5cm,height=16.5cm}}
\noindent
{\bf Figure 6.} The same as Figure 4, for a model
with $(\Omega_0, \Gamma, \sigmaba)=(0.2,0.2,1.07)$.
Crosses show the same results for SO clusters.
\end{figure}

\begin{figure}
\centerline{
\psfig{figure=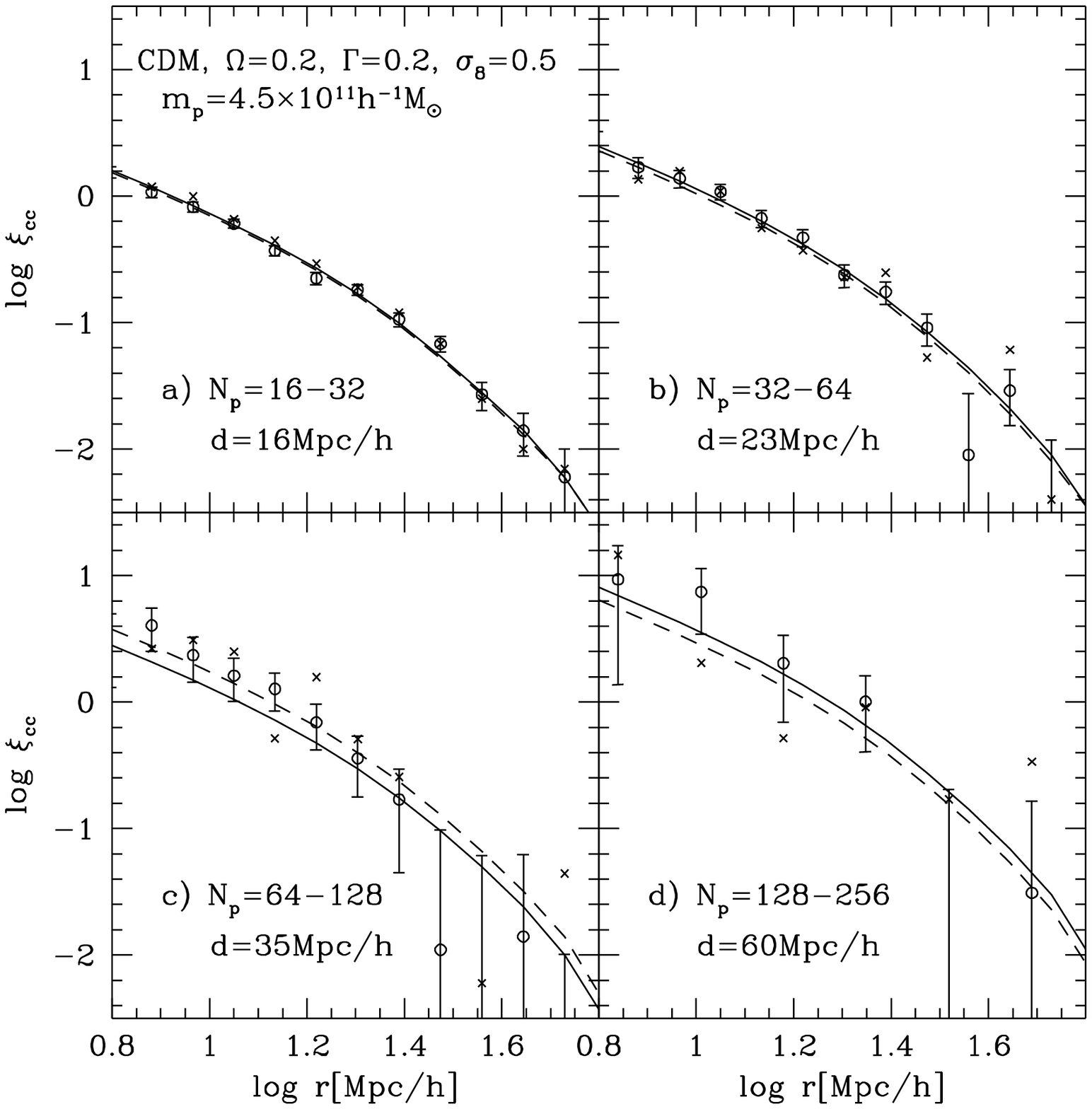,width=14.5cm,height=16.5cm}}
\noindent
{\bf Figure 7.} The same as Figure 4, for a model
with $(\Omega_0, \Gamma, \sigmaba)=(0.2, 0.2, 0.5)$.
Crosses show the same results for SO clusters.
\end{figure}

\begin{figure}
\centerline{
\psfig{figure=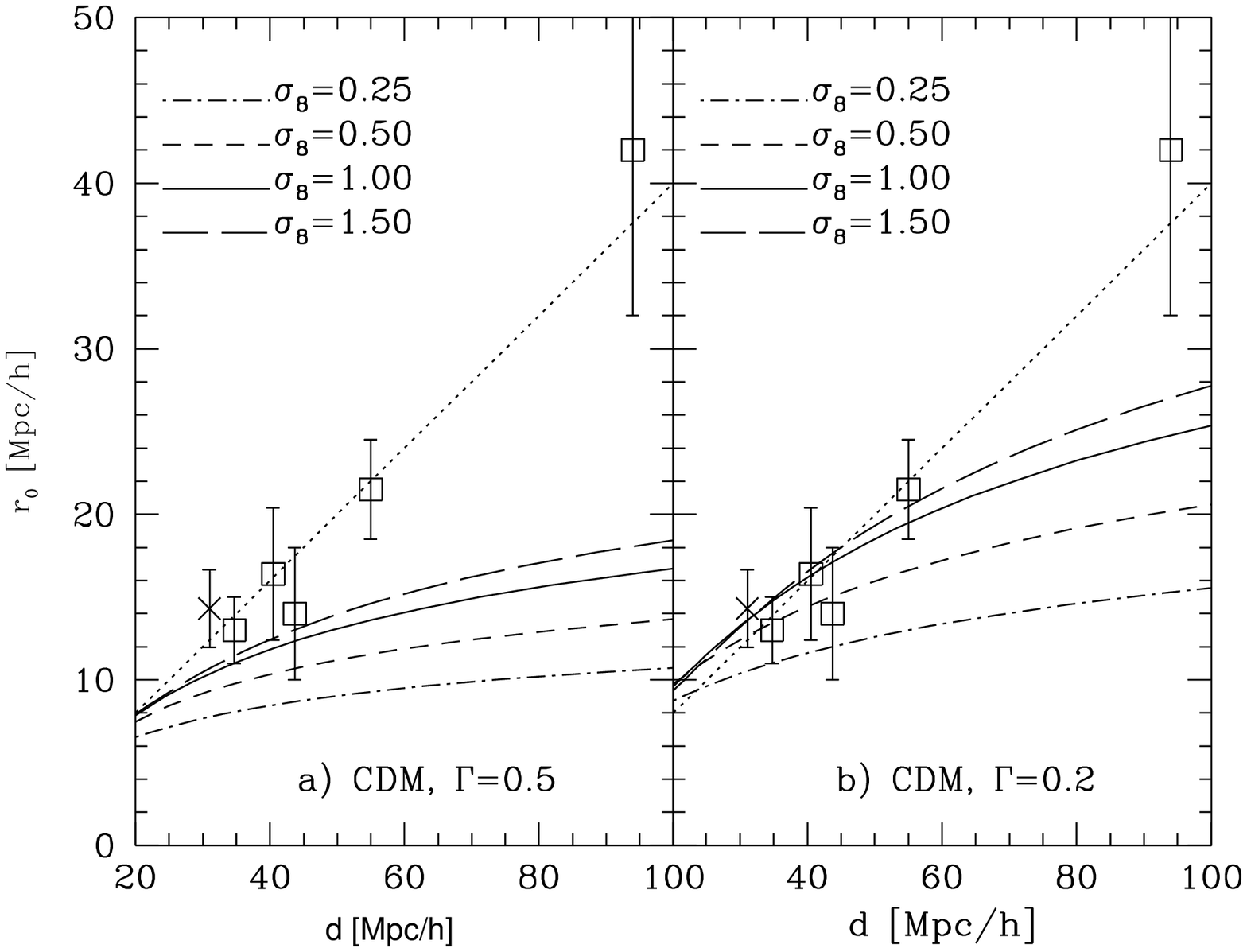,width=14.5cm,height=16.5cm}}
\noindent
{\bf Figure 8.} Correlation length $r_0$ as 
a function of the mean intercluster separation $d$,
for two sets of models with $\Gamma =0.5$ (a) and
$\Gamma=0.2$ (b). For a given $\Gamma$, the $d$-$r_0$
relation does not depend on $\Omega_0$, but only on $\sigmaba$.
The data points (squares) are observational results
compiled by Bahcall \& Cen (1992), and the error bars are 1$\sigma$. 
In addition, 
the result of a recent determination by 
 Dalton et al. (1994) for APM clusters is also plotted (the cross)
together with its 2$\sigma$ error bar.
The dotted lines
show the relation $r_0=0.4d$.
\end{figure}
\end{document}